\documentclass{PoS}

\usepackage{amsmath,amssymb}
\usepackage{graphicx}
\usepackage{braket}

\newcommand{\ii}{\ensuremath{\mathrm{i}}}

\title{Massive gluon propagator\\ at zero and finite temperature}

\ShortTitle{Massive gluon at zero and finite $T$}

\author{Attilio Cucchieri$^{a,b}$, David Dudal$^b$,
        \speaker{Tereza Mendes}$^a$ and Nele Vandersickel$^b$\\
\llap{$^a$}Instituto de F\'\i sica de S\~ao Carlos, 
           Universidade de S\~ao Paulo,\\
           Caixa Postal 369, 13560-970 S\~ao Carlos, SP, Brazil\\
\llap{$^b$}Ghent University, Department of Physics and Astronomy,\\
           Krijgslaan 281-S9, 9000 Gent, Belgium\\
E-mail: \email{attilio@ifsc.usp.br},\email{david.dudal@ugent.be},\\
        \hskip 12mm 
        \email{mendes@ifsc.usp.br},\email{nele.vandersickel@ugent.be}}

\abstract{
We report on our study of the infrared gluon propagator for SU(2)
lattice gauge theory using large lattice volumes. The observed massive
behavior is discussed from the point of view of analytic predictions for
the zero-temperature case. Such a behavior is still present as the
temperature is switched on, but manifests itself differently in the
electric and magnetic channels.
          }

\FullConference{International Workshop on QCD Green's Functions, Confinement and Phenomenology,\\
		September 05-09, 2011\\
		Trento Italy}

\begin{document}

\section{Introduction}

The massive infrared behavior of the gluon propagator $D(p^2)$ in Landau 
gauge (see e.g.\ \cite{Aguilar:2008xm}) has been distinctively observed 
in lattice simulations using very large volumes a few years ago
\cite{Bogolubsky:2007ud,Cucchieri:2007md,Sternbeck:2007ug}, 
for SU(2) and SU(3) pure gauge theory (see \cite{Cucchieri:2010xr}
for a review). This behavior has been
recently analyzed in terms of an effective running mass in
\cite{Aguilar:2011yb} (see also \cite{Oliveira:2010xc})
and it has been associated to the dimension-two
condensates in the so-called refined Gribov-Zwanziger (RGZ) framework
(see e.g.\ \cite{Dudal:2008sp}) for the SU(3) case in \cite{Dudal:2010tf}.
Here we present the results of our fits to the SU(2) gluon propagator,
which can be associated to the RGZ formula. More precisely, we have
performed systematic fits to our data in the whole range of available 
momenta (in the infrared region) using the so-called Gribov-Stingl form 
\cite{Stingl:1985hx,Stingl:1994nk}
for modeling the massive behavior of the gluon propagator.
This form is a generalization of the Gribov propagator 
\cite{Gribov:1977wm,Zwanziger:1991gz} --- which is based on a pair 
of purely imaginary complex-conjugate poles --- to include pairs of 
complex-conjugate poles with a nonzero real part, as well as a possible 
real pole.
We have tested several rational forms of this type for $D(p^2)$, and 
found that a four-parameter expression (in which one of the parameters 
is a global normalization) gives the best quantitative description of 
the data in the 4d case. In the 3d case we need five parameters, one of 
which again serves as a global normalization. 
The behavior in the two cases is associated respectively to the simplified 
and to the general RGZ formulas for $D(p^2)$.
In two dimensions, on the contrary,
the use of rational forms is not sufficient to describe the data. This
case will not be considered here. A more detailed discussion of these
zero-temperature fits has been presented recently in \cite{Cucchieri:2011ig}.

At finite temperature, a similar massive behavior is observed for both
the longitudinal (electric) and the transverse (magnetic) gluon propagator
in the infrared limit \cite{Cucchieri:2007ta,Fischer:2010fx,Bornyakov:2010nc,Cucchieri:2011ga,Bornyakov:2011jm,Cucchieri:2011di,Aouane:2011fv,Maas:2011ez}.
In this case, we have used a modified Gribov-Stingl form to describe
our SU(2) data and to define electric and magnetic screening masses.
A recent update on our results can be found in \cite{Lat11}.
This study is still preliminary. 

In the following, we review the features of the RGZ framework
in Section \ref{RGZ}, then present our fit results in Section \ref{fits}.
The finite-temperature case is considered in Section \ref{finiteT},
followed by our conclusions and the bibliography.

%%%%%%%%%%%%%%%%%%%%%%%%%%%%%%%%%%%%%%%%%%%%%%%%%%%%%%%%%%%%%%%%%%%%%%%%

\section{The Refined Gribov-Zwanziger Framework}
\label{RGZ}

The refined Gribov-Zwanziger framework (RGZ) differs from the scenario 
originally proposed by
Gribov \cite{Gribov:1977wm} and Zwanziger \cite{Zwanziger:1991gz}
through the introduction of dimension-two
condensates
\cite{Dudal:2008sp,Dudal:2008rm,Vandersickel:2011ye,Vandersickel:2011zc,Dudal:2011gd}. 
In the most general case, four different condensates are 
considered, i.e.\
\begin{align}
\braket{A_\mu^a A_\mu^a}                                & \to -m^2  &
\Braket{\overline{\varphi}^a_i \varphi^a_{i}}           & \to M^2  &
\Braket{\varphi^a_i \varphi^a_{i}}                      & \to \rho &
\Braket{\overline{\varphi}^a_i \overline \varphi^a_{i}} & \to \rho^\dagger \; ,
\label{eq:condensates}
\end{align}
where we have listed the dynamical mass associated to each condensate.
Note that the condensate $\,-m^2$ is directly related to the gluon condensate
$\langle g^2 A^2 \rangle$.
In the presence of the four condensates above, the original infrared 
suppressed gluon propagator in \cite{Gribov:1977wm,Zwanziger:1991gz}
is modified as
\footnotesize
\begin{equation}
D(p^2) \; = \; \frac{p^4 + 2 M^2 p^2 + M^4 - (\rho_1^2 + \rho_2^2)}{
          p^6 + p^4 \left (m^2 + 2 M^2 \right) + p^2 \left[2 m^2 M^2 + M^4
          + \lambda^4 - \left( \rho_1^2 + \rho_2^2 \right) \right]
          + m^2 \left[ M^4 - \left( \rho_1^2 + \rho_2^2
          \right) \right] + \lambda^4 \left( M^2 - \rho_1 \right)} \; .
\label{refinedgluonprop}
\end{equation}
\normalsize
where the condensates $m^2$, $M^2$, $\rho$ are described above and
$\lambda^4$ is related to the Gribov parameter $\gamma$ through
$\lambda^4 = 2 g^2 N_c \gamma^4$. Also, we have set
\begin{eqnarray}
\rho &=& \rho_1 + \ii \rho_2 \nonumber\\
\rho^\dagger &=& \rho_1 - \ii \rho_2\,.
\end{eqnarray}
It is interesting to notice that this propagator gets simplified if
$\rho = \rho^\dagger = \rho_1$ (i.e.\ $\rho_2=0$), which corresponds
to the equality $\Braket{ \, \overline{\varphi} \overline{\varphi} \, }
= \Braket{ \, \varphi \varphi \,}$ from (\ref{eq:condensates}).
Indeed, in this case one can factorize the quantity $p^2 + M^2 - \rho_1\,$
in the numerator and in the denominator of the above formula, obtaining
\begin{equation}\label{prop}
D(p^2) \; = \; \frac{p^2 + M^2 + \rho_1}{p^4 + p^2 \left( M^2 + m^2 +
      \rho_1 \right) + m^2 \left( M^2 + \rho_1 \right) +  \lambda^4} \;.
\end{equation}

Note that both Eqs.\ \eqref{refinedgluonprop} and \eqref{prop} can be
decomposed as sums of propagators of the type $ \alpha / (p^2+\omega^2)$.
In particular, we can write Eq.\ \eqref{refinedgluonprop} as
\begin{equation}\label{gluonpropsimp}
D(p^2) \; = \; \frac{\alpha}{p^2+\omega_1^2} \,+\, \frac{\beta}{p^2+\omega_2^2}
\,+\, \frac{\gamma}{p^2+\omega_3^2} \;\,.
\end{equation}
To this end, we only need to solve the cubic equation
\begin{eqnarray}
x^3 + x^2 \left (m^2 + 2 M^2 \right) + x \left[2 m^2 M^2 + M^4 + \lambda^4 -
\left( \rho_1^2 + \rho_2^2 \right) \right] 
+ m^2 \left[ M^4 - \left( \rho_1^2 + \rho_2^2 \right) \right] & &
\nonumber \\[1mm]
+ \lambda^4 \left( M^2 - \rho_1 \right) & = & 0 \; , \quad
\end{eqnarray}
obtained by setting $p^2 = x\,$ in the denominator of
Eq.\ \eqref{refinedgluonprop}, and to find its three roots
$\omega_1^2$, $\omega_2 ^2$ and $\omega_3^2$. At the same time, the gluon
propagator in Eq.\ \eqref{prop} can be written as
\begin{equation}\label{4D3}
D(p^2) \; = \; \frac{\alpha_+}{p^2+\omega_{+}^2} \,+\,
\frac{\alpha_-}{p^2+\omega_{-}^2} \;,
\end{equation}
where we expect to have $\alpha_- = \alpha_+^*$ if
$\,\omega_{-}^2 = (\omega_{+}^2)^*$, i.e.\ if $\,\omega_{+}^2$ and
$\omega_{-}^2$ are complex conjugates.
Here, $\omega_{\pm}^2$ are the roots of the quadratic equation
\begin{equation}
x^2 + x \left( M^2 + m^2 + \rho_1 \right) + m^2 \left( M^2 + \rho_1 \right) +
\lambda^4 \; = \; 0 \; ,
\end{equation}
obtained by setting $p^2 = x$ in the denominator of Eq.\ \eqref{prop}.
Clearly, one finds complex-conjugate poles if
$\,|M^2 - m^2 + \rho_1| \,<\, 2\lambda^2$.

\vskip 3mm
Let us remark that rational forms such as \eqref{refinedgluonprop} and
\eqref{prop} for the gluon propagator were considered by Stingl
\cite{Stingl:1985hx,Stingl:1994nk}, as a way of accounting for
nonperturbative effects in an extended perturbative approach to Euclidean QCD.
More precisely, in his treatment, one expresses the proper vertices of
the theory as an iterative sequence of functions yielding a self-consistent
solution to the Dyson-Schwinger equations.
In particular, for the gluon propagator, this sequence is written
[see Eq.\ (2.10) in Ref.\ \cite{Stingl:1994nk}]
in terms of ratios of polynomials in the variable $p^2$, of degree $r$ in
the numerator and $r+1$ in the denominator, with $r = 0, 1, 2, \ldots\;$.
This functional form is then related, via operator the product expansion, to
the possible existence of vacuum condensates of dimension $2n$, with $n\geq 1$.
At the same time, the associated complex-conjugate poles are
interpreted as short-lived elementary excitations of the gluon field
\cite{Stingl:1985hx,Zwanziger:1991gz,Stingl:1994nk}.
By comparison, in the RGZ framework, one proposes specific forms for
the dimension-two condensates --- related to the auxiliary fields of
the GZ action --- and then obtains (at tree level) the rational
functions in Eqs.\ (\ref{refinedgluonprop}) and (\ref{prop}),
which correspond respectively to cases with $r = 3$ and 2 in Stingl's
iterative sequence.

%%%%%%%%%%%%%%%%%%%%%%%%%%%%%%%%%%%%%%%%%%%%%%%%%%%%%%%%%%%%%%%%%%%%%%%%%%%

\section{Zero-temperature results}
\label{fits}

We analyze data for the SU(2) Landau-gauge gluon propagator, produced
in 2007 and already discussed in
\cite{Cucchieri:2007md,Cucchieri:2007rg,Cucchieri:2008fc,Cucchieri:2010xr},
but not systematically fitted until recently. Our run parameters and
lattice setup are described in \cite{Cucchieri:2011ig}.
We note that the lattice spacing $a$ is set by using the 4d SU(3) value 
for the string tension, as described in \cite{Cucchieri:2003di} and
\cite{Bloch:2003sk} respectively for d = 3 and 4. 
All our runs are in the scaling region.
Possible systematic effects due to Gribov copies
as well as unquenching effects are {\em not} considered here.
Finite-volume effects, on the other hand, are well under control
and our largest lattice volumes can be already considered as infinite.
We notice that, in order to reduce discretization errors due to
the breaking of rotational symmetry, we have considered several
configurations for the momentum components $p_{\mu}$ and used
the improved momentum definition in \cite{Ma:1999kn}, which does not 
affect the value of $p^2$ in the IR limit, but modifies its value 
significantly for large momenta.
We have checked that the use of improved momenta helps to obtain a 
better fit to the data in both the 4d and the 3d cases.

Values of physical parameters (i.e.\ the condensates and poles introduced
in the previous section) are extracted from the data at the largest 
lattices, with lattice volume $128^4$ in 4d and $320^3$ in 3d, and 
lattice spacing respectively of 0.210 fm and 0.268 fm. This corresponds
to physical volumes of about $(27~\mbox{fm})^4$ and $(85~\mbox{fm})^3$,
or equivalently smallest momenta of about 46 MeV and 14 MeV, respectively
in 4d and 3d.

Our results are summarized below.
We refer to \cite{Cucchieri:2011ig} for a more complete analysis.
We remark that the shown data for $D(p^2)$ are {\em not} normalized and
that a renormalization condition at a given scale would
correspond to a rescaling of the overall factor $C$ in the fitting
forms considered below. The condensates and the poles, on the other hand,
are not affected by such a renormalization.
Also note that, since our largest momentum is of the order of 
$4 ~\mbox{GeV}$, ultraviolet logarithmic corrections are not important to
describe the lattice data and they are not included in the fitting functions
proposed here. This also avoids the problem of having to regularize
the corresponding Landau pole by hand.

%%%%%%%%%%%%%%%%%%%%%%%%%%%%%%%%%%%%%%%%%%%%%%%%%%%%%%%%%%%%%%%%%%%%%%%%%%%%

\begin{figure}
\begin{center}
\includegraphics[width=.75\textwidth]{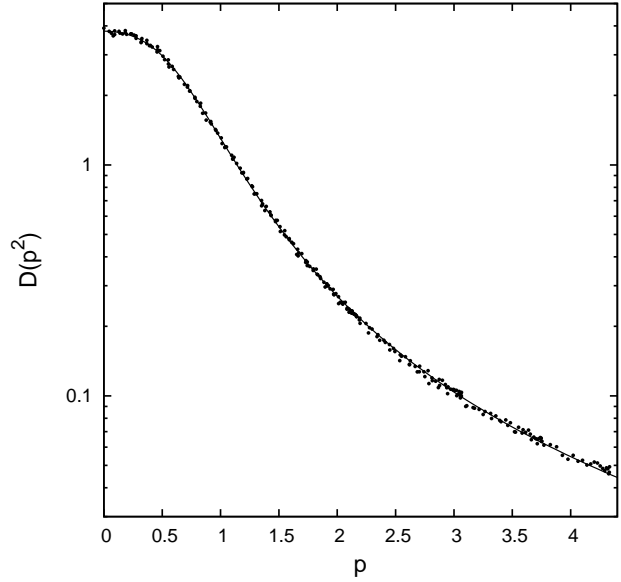}
\caption{Plot of the 4d gluon propagator $D(p^2)$ (in $\mbox{GeV}^{-2}$) as
a function of the (improved) momentum $p$ (in GeV) for the lattice volume
$V = 128^4$. We also show the fitting function $f_{1}(p^2)$.
Note the logarithmic scale on the $y$ axis.}
\label{fig:4dgl}
\end{center}
\end{figure}

\vskip 3mm
In the 4d case, our best fit is obtained for a four-parameter fitting
function of the simplest Gribov-Stingl form
\begin{equation}
f_{1}(p^2) \; = \; C \, \frac{p^2 + s}{p^4 + u^2 \, p^2 + t^2} \; ,
\label{eq:f4dgluon}
\end{equation}
which corresponds to the simplified RGZ propagator in Eq.\ (\ref{prop}),
modulo the global rescaling factor $C$.
The results of the fit for all lattice volumes, using improved momenta, 
are reported in Table \ref{tab:gluon4d-im}. 
The good quality of the fit is seen by comparing it to the data, 
as shown (for our largest lattice) in Fig.\ \ref{fig:4dgl}.
Let us stress that we are fitting the whole momentum range available and
that, for the largest lattice volume, we have 257 data points.
\begin{table}
\begin{center}
\begin{tabular}{cccccc}
\hline
\hline
  $ V $  &   $  C  $      &   $ u (\mbox{GeV})$   &  $t (\mbox{GeV}^2)$   & $ s (\mbox{GeV}^2)$   & $\chi^2$/d.o.f.\ \\
\hline
 $48^4$  &  0.791 (0.007) &  0.755 (0.027) &  0.707 (0.013) &  2.419 (0.119) &  2.09 \\
\hline
 $56^4$  &  0.801 (0.006) &  0.734 (0.023) &  0.696 (0.012) &  2.305 (0.100) &  1.92 \\
\hline
 $64^4$  &  0.791 (0.007) &  0.760 (0.024) &  0.710 (0.012) &  2.425 (0.108) &  2.35 \\
\hline
 $80^4$  &  0.785 (0.005) &  0.734 (0.019) &  0.708 (0.009) &  2.404 (0.084) &  2.04 \\
\hline
 $96^4$  &  0.795 (0.004) &  0.717 (0.016) &  0.694 (0.008) &  2.291 (0.068) &  1.66 \\
\hline
 $128^4$ &  0.784 (0.005) &  0.768 (0.017) &  0.720 (0.009) &  2.508 (0.078) &  1.63 \\
\hline
\hline
\end{tabular}
\end{center}
\caption{
Fits of the gluon-propagator data in the 4d case, for different lattice
volumes, using the fitting function $f_1(p^2)$ and improved momenta.
We report, besides the value of the fit parameters, the $\chi^2$/d.o.f.\
obtained in each case. The whole range of momenta was considered for the fit.
Errors shown in parentheses correspond to one standard deviation.
}
\label{tab:gluon4d-im}
\end{table}

We mention that a test of the more general form
of the propagator [given in Eq.\ (\ref{refinedgluonprop})]
considering a six-parameter fitting function leads to an unstable fit, in 
which most of the parameters are determined with very large errors,
suggesting that the function has too many (redundant) parameters.
We then reduced the number of parameters by one and introduced the form
\begin{equation}
f_{4}(p^2) \;=\; C \,\frac{\left(p^2 + s\right)\,\left(p^2 + 1\right)}{
\left(p^4 + u^2 \, p^2 + t^2\right)\,\left(p^2 + k\right)} \;=\; C \,
\frac{p^4 + (s+1) p^2 + s}{p^6 + (k+u^2) p^4 + (k u^2+t^2) p^2 + k t^2} \;,
\label{eq:fDvrgzsimple}
\end{equation}
which is written as a simple generalization of
$f_1(p^2)$ in Eq.\ (\ref{eq:f4dgluon}).
In this case the fits look reasonable, but the errors are larger and 
the $\chi^2$/d.o.f.\ is not better for the five-parameter fit compared
to the four-parameter fit, indicating that the latter is more stable.
Also, the fits results suggest a very small (and imaginary) value for 
$\rho_2$, implying that $\rho$ is real and thus supporting the simpler 
form in Eq.\ (\ref{prop}), fitted above using the function $f_1(p^2)$.

In order to extract the value of the condensates described in Section
\ref{RGZ} above, we thus consider only the fit results for $f_1(p^2)$ and 
the volume $V=128^4$ (using improved momenta), reported in the last row of
Table \ref{tab:gluon4d-im} and plotted in Fig.\ \ref{fig:4dgl}.
By setting $f_{1}(p^2)$ equal to the RGZ propagator in Eq.\ (\ref{prop})
(modulo the global factor $C$), we find for the condensates the values 
reported in Table \ref{tab:gluon4d-param}. 
We see that $\,|M^2-m^2+\rho_1| < 2\lambda^2$, justifying our
expectation that the propagator
may be decomposed in terms of a pair of complex-conjugate poles.
We can thus write [see Eq.\ (\ref{4D3})]
\begin{equation}
f_{2}(p^2) \;=\; \frac{\alpha_+}{p^2+\omega_+^2} \,+\,
\frac{\alpha_-}{p^2 + \omega_-^2} \;=\;
\frac{2 a \,p^2 \,+\, 2(a v + b w)}{p^4 \,+\, 2 v \,p^2 \,+\, v^2 + w^2} \; ,
\label{eq:fpoles4d}
\end{equation}
with $\alpha_{\pm} = a \pm i b $ and $\omega_{\pm}^2 = v \pm i w $.
The results for the parameters $a, b, v$ and $w$ are also shown in 
Table \ref{tab:gluon4d-param}.
We note that the errors (given in parentheses) correspond to one
standard deviation and were evaluated in three different ways:
by propagation of error, by a Monte Carlo error analysis and by
a bootstrap analysis. We refer to \cite{Cucchieri:2011ig} for
details of these procedures.
Clearly, all results obtained agree within errors.
We see that the poles are complex conjugates whose imaginary part is more than
twice their real part. We recall that a Gribov propagator would have a null
real part.
\begin{table}
\begin{center}
\begin{tabular}{cccc}
\hline
\hline
parameter $\;$&$\;$ propagation of error $\;$ & $\;$ Monte Carlo analysis
$\;$ & $\;$ bootstrap analysis \\
\hline
$M^2+\rho_1\,(\mbox{GeV}^2)$     & $2.51(8)$  & $2.51(8)$  & $2.3(3)$  \\
\hline
$m^2\,(\mbox{GeV}^2)$       & $-1.92(9)$ & $-1.92(9)$ & $-1.7(2)$ \\
\hline
$\lambda^4\,(\mbox{GeV}^4)$ & $5.3(9)$   & $5.3(4)$   &  $4.5(9)$ \\
\hline
$a$                   & 0.392(3)  &  0.392(2) &  0.38(1)  \\
\hline
$b$                   & 1.32(7)   &  1.32(5)  &  1.20(7)  \\
\hline
$v \,(\mbox{GeV}^2)$  & 0.29(2)   &  0.29(2)  &  0.29(3)  \\
\hline
$w \,(\mbox{GeV}^2)$  & 0.66(2)   &  0.66(1)  &  0.64(2)  \\
\hline
\hline
\end{tabular}
\end{center}
\caption{
Estimates of the parameters of the simplified RGZ gluon propagator
in Eq.\ (2.4) and of the function $f_{2}(p^2)$, obtained from fits 
to the equivalent form $f_1(p^2)$.
Errors are calculated using propagation of error, a Monte Carlo analysis
and a bootstrap analysis. 
In all cases we considered the volume $V = 128^4$ and improved momenta.
}
\label{tab:gluon4d-param}
\end{table}

Let us mention that the values obtained here for $M^2+\rho_1$, $m^2$ and
$\lambda^4$ are in good quantitative agreement with the corresponding
values --- respectively indicated with $M^2$, $m^2$ and $ 2 g^2 N \gamma^4 $
--- reported in Ref.\ \cite{Dudal:2010tf} for the SU(3) case.\footnote{For
comparison with our values in Table \ref{tab:gluon4d-param}, the
SU(3) condensates from \cite{Dudal:2010tf} are respectively
2.15(13) GeV$^2$, $-1.81(14)$ GeV$^2$ and 4.16(38) GeV$^4$.}
Also, as remarked above, the condensate $m^2$ may be used to obtain a value
for the gluon condensate $\langle g^2 A^2 \rangle$, through the relation
(see e.g.\ \cite{Dudal:2010tf})
\begin{equation}
\langle g^2 A^2 \rangle\;=\; - \frac{9}{13} \,\frac{N_c^2-1}{N_c} \, m^2\,.
\end{equation}
In our case, the value $m^2=-1.92(9)$ from Table \ref{tab:gluon4d-param}
(using propagation of error) yields
$\langle g^2 A^2 \rangle = 1.99(9)$ GeV$^2$.

%%%%%%%%%%%%%%%%%%%%%%%%%%%%%%%%%%%%%%%%%%%%%%%%%%%%%%%%%%%%%%%%%%%

\begin{figure}[t]
\begin{center}
\includegraphics[width=.75\textwidth]{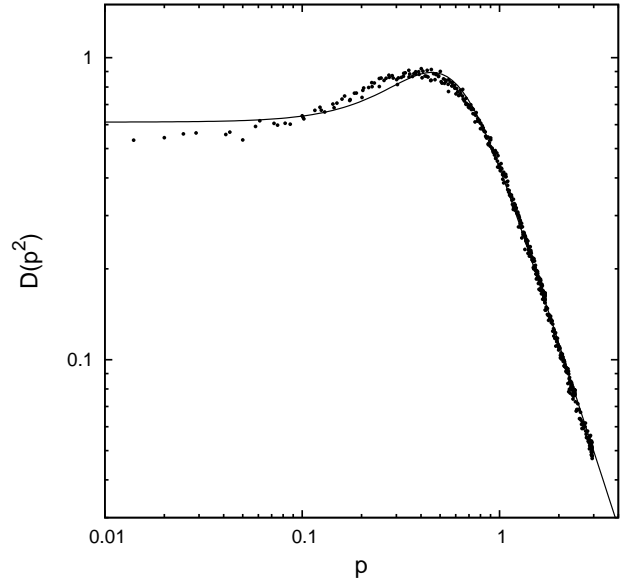}
\caption{Plot of the 3d gluon propagator $D(p^2)$ (in $\mbox{GeV}^{-1}$) as
a function of the (improved) momentum $p$ (in GeV) for the lattice volume
$V = 320^3$. We also show the fitting function $f_{1}(p^2)$.
Note the logarithmic scale on both axes.}
\label{fig:3dgl-rgz}
\end{center}
\end{figure}

\begin{figure}
\begin{center}
\includegraphics[width=.75\textwidth]{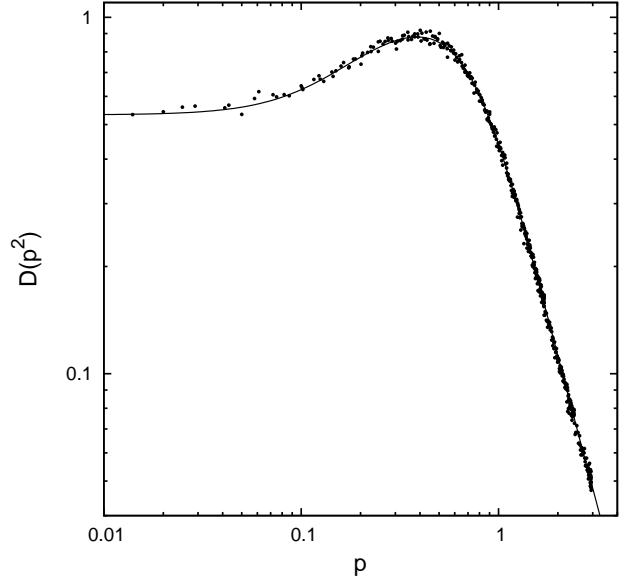}
\caption{Plot of the 3d gluon propagator $D(p^2)$ (in $\mbox{GeV}^{-1}$) as
a function of the (improved) momentum $p$ (in GeV) for the lattice volume
$V = 320^3$. We also show the fitting function $f_{4}(p^2)$.
Note the logarithmic scale on both axes.}
\label{fig:3dgl}
\end{center}
\end{figure}

\vskip 3mm
In the 3d case the simplified fitting form $f_{1}(p^2)$ in
Eq.\ (\ref{eq:f4dgluon}) is not able to describe well the lattice data.
Indeed, even using improved momenta,
the $\chi^2$/d.o.f.\ values obtained are quite large. Moreover, as can
be seen in Fig.\ \ref{fig:3dgl-rgz}, the fit clearly fails in the IR
region.\footnote{In order to highlight the results at small momenta, here
and in Fig.\ \ref{fig:3dgl} we present the plot with a logarithmic scale
on both axes.}
The situation improves by considering the (five-parameter) fitting function
$f_4(p^2)$ in Eq.\ (\ref{eq:fDvrgzsimple}) above, as can be seen from the
results reported in Table \ref{tab:gluon3d-im},
obtained using improved momenta.
\begin{table}
\begin{center}
\begin{tabular}{cccccc}
\hline
\hline
 $ V $ &  $C (\mbox{GeV})$    &   $ u (\mbox{GeV})$   &  $t (\mbox{GeV}^2)$   &
 $ s (\mbox{GeV}^2)$   & $ k (\mbox{GeV}^2)$  \\
\hline
$140^3$ &  0.407 (0.001) &  0.654 (0.008) &  0.623 (0.004) &  0.022 (0.002) &  0.041 (0.003) \\
\hline
$200^3$ &  0.407 (0.001) &  0.655 (0.007) &  0.623 (0.004) &  0.024 (0.002) &  0.043 (0.003) \\
\hline
$240^3$ &  0.408 (0.001) &  0.662 (0.007) &  0.620 (0.004) &  0.025 (0.002) &  0.047 (0.003) \\
\hline
$320^3$ &  0.408 (0.001) &  0.656 (0.008) &  0.619 (0.005) &  0.023 (0.002) &  0.046 (0.004) \\
\hline
\hline
\end{tabular}
\end{center}
\caption{
Fits of the gluon-propagator data in the 3d case, for different lattice
volumes, using the fitting function $f_4(p^2)$ and improved momenta.
The $\chi^2$/d.o.f.\ is about 1 for the lattice volume $320^3$.
The whole range of momenta was considered for the fit.
Errors shown in parentheses correspond to one standard deviation.
}
\label{tab:gluon3d-im}
\end{table}
Let us mention that we have also tried a six-parameter fit to a more 
general function, obtained by substituting $(p^2 + 1)$ in the numerator
of $f_4(p^2)$ in Eq.\ (\ref{eq:fDvrgzsimple}) by $(p^2 + l)$.
In this case we obtain a good fit (with $\chi^2$/d.o.f.\ around 1),
with $l \approx 1$ and values of the other parameters that are consistent 
with the ones in Table \ref{tab:gluon3d-im}, indicating that the latter 
fit is preferable.

In order to evaluate the condensates of the RGZ model, we thus consider only
the results from the fit using $f_4(p^2)$, given for the lattice size $N=320$
in the last row of Table \ref{tab:gluon3d-im} and plotted in
Fig.\ \ref{fig:3dgl}.
By setting $f_{4}(p^2)$ [see Eq.\ \ref{eq:fDvrgzsimple})] equal to the RGZ
propagator (\ref{refinedgluonprop}) modulo the global factor $C$, we find
the values for the condensates in Table \ref{tab:gluon3d-param}.
Note that, using this fitting form, we are able to evaluate $M^2$, $\rho_1$
and $|\rho|$ (and therefore $\rho_2$) separately.
In this case, we can see that $\rho_2 \neq 0$ and $\rho$ is indeed a complex
quantity. This is consistent with the fact that the (four-parameter) fit to
the simplified form $f_1(p^2)$ fails, as seen above.
Finally, we decompose the propagator as in Eq.\ (\ref{gluonpropsimp}) with
$\beta = a + i b $, $\gamma = a - i b $, $\omega_2^2 = v + i w $ e
$\omega_3^2 = v - i w $, i.e.\ we consider the function
\begin{equation}
f_{6}(p^2) \; = \;
\frac{\alpha}{p^2 + \omega_1^2} \, + \,
\frac{2 a \, p^2 \,+\, 2(a v + b w)}{p^4 \,+\, 2 v\, p^2 \,+\, v^2 + w^2} \,.
\label{eq:fpoles3d}
\end{equation}
The corresponding results are also reported\footnote{Clearly, we have 
$\omega_1^2 = k$ from $f_4(p^2)$.} in Table \ref{tab:gluon3d-param}.
Also in this case we have performed the error analysis in three 
different ways: propagation of error, Monte Carlo error analysis and
bootstrap analysis (see \cite{Cucchieri:2011ig} for details).
Note that the imaginary part $w$ of the complex-conjugate poles is more 
than twice the value of their real part $v$, as in the 4d case.
Note also that the mass $\omega_1$ and the residue $\alpha$
associated with the real pole are very small. 
Moreover, $\alpha$ is negative, which may
be associated with violation of reflection positivity, indicating that
this mass cannot correspond to a physical degree of freedom.
\begin{table}
\begin{center}
\begin{tabular}{cccc}
\hline
\hline
 parameter  $\;$&$\;$ propagation of error $\;$ & $\;$ Monte Carlo analysis $\;$ & $\;$ bootstrap analysis \\
\hline
 $M^2 \,(\mbox{GeV}^2)$  & 0.512 (1) & 0.512 (1) & 0.513 (1) \\
\hline
 $m^2 \,(\mbox{GeV}^2)$  & $-0.55 (1)$ & $-0.55 (1)$ & $-0.52 (2)$ \\
\hline
 $\lambda^4 \,(\mbox{GeV}^4)$ & 0.94 (1) & 0.94 (1) & 0.91 (3) \\
\hline
 $\rho_1 \,(\mbox{GeV}^2)$ & 0.479 (2) & 0.479 (2) & 0.477 (2) \\
\hline
 $\rho_2 \,(\mbox{GeV}^2)$ & 0.09 (1) & 0.094 (9) & 0.100 (6)    \\
\hline
 $\alpha \,(\mbox{GeV})$ & $-0.024 (5)$ & $-0.024 (5)$ & $-0.029 (4)$  \\
\hline
 $\omega_1^2 \,(\mbox{GeV}^2)$ & 0.046 (4) & 0.046 (4) & 0.046 (4) \\
\hline
 $a \,(\mbox{GeV})$ & 0.216 (3) & 0.216 (2) & 0.220 (4)  \\
\hline
 $b \,(\mbox{GeV})$ & 0.27 (5) & 0.271 (3) & 0.275 (3) \\
\hline
 $v \,(\mbox{GeV}^2)$ & 0.215 (5) & 0.215 (5) & 0.23 (1) \\
\hline
 $w \,(\mbox{GeV}^2)$ & 0.580 (6) & 0.580 (6) & 0.57 (1) \\
\hline
\hline
\end{tabular}
\end{center}
\caption{
Estimates of the parameters of the general RGZ propagator in Eq.\ (2.1)
and of the function $f_{6}(p^2)$, obtained from fits to the equivalent 
form $f_4(p^2)$.
Errors are calculated using propagation of error, a Monte Carlo analysis
and a bootstrap analysis. 
In all cases we considered the volume $V = 320^3$ and improved momenta.
}
\label{tab:gluon3d-param}
\end{table}

%%%%%%%%%%%%%%%%%%%%%%%%%%%%%%%%%%%%%%%%%%%%%%%%%%%%%%%%%%%%%%%%%%%%%%%%%%

\section{Finite temperature}
\label{finiteT}

We have used a modified Gribov-Stingl expression to fit our infrared
data for finite-temperature Landau-gauge SU(2) gluon propagators
(in 3$+$1 dimensions) reported in \cite{Cucchieri:2011ga,Cucchieri:2011di,Lat11}.
For both the longitudinal (electric) propagator $D_L(p^2)$ and for 
the transverse (magnetic) propagator $D_T(p^2)$,
we consider the five-parameter fitting form\footnote{Note that, as in 
the previous section, the global constant $C$ is fixed (for given values 
of $a$, $b$, $d$, $\eta$) by the renormalization condition,
so that there are only four free parameters in (\ref{GSform}).}
\begin{equation}
D_{L,T}(p^2) \;=\;
C\,\frac{1\,+\,d\,p^{2 \eta}}
{(p^2 + a)^2 \,+\, b^2}\,.
\label{GSform}
\end{equation}
This form allows for two (complex-conjugate) poles, with
masses $\,m^2 \;=\; a\,\pm\, i b$, where $m \;=\; m_R \,+\, i m_I$.
The mass $m$ thus depends only on $a$, $b$ and not on the normalization $C$.
The parameter $\eta$ should be 1 if the fitting form also describes
the large-momenta region (from our infrared data we get $\eta\neq 1$).
Recall that at high temperatures one usually defines the electric screening
mass as the scale determining the exponential decrease of the real-space
propagator at large distances, which is equivalent to $\,D_L(0)^{-1/2}$
in the case of a real pole.
We therefore expect to observe $\,m_I\to 0\,$ (i.e.\ $\,b\to 0\,$)
for the longitudinal gluon propagator at high temperature.
Note that, if the propagator has the above form (with nonzero $b$), then
the screening mass defined by $\,D_L(0)^{-1/2} \,=\, \sqrt{(a^2+b^2)/C}\;\,$
mixes the complex and imaginary masses $\,m_R$ and $m_I\,$ and depends
on the (a priori arbitrary) normalization $C$.

We generally find good fits to the modified Gribov-Stingl form 
above (including the full range of momenta), with
nonzero real and imaginary parts of the pole masses in all cases.
For the transverse propagator $D_T(p^2)$, the masses $m_R$ and $m_I$
are of comparable size (around 0.6 and 0.4 GeV respectively).
The same holds for $D_L(p^2)$, but in this case the relative size of
the imaginary mass seems to decrease with increasing temperature.
A detailed discussion of the associated masses $m_R$, $m_I$ is
postponed to a forthcoming study \cite{FT}, as we are presently
considering variants of the above fitting form inspired by
the zero-temperature forms considered in the previous sections.

\begin{figure}
\hspace*{-1.5cm}
\includegraphics[height=7.truecm]{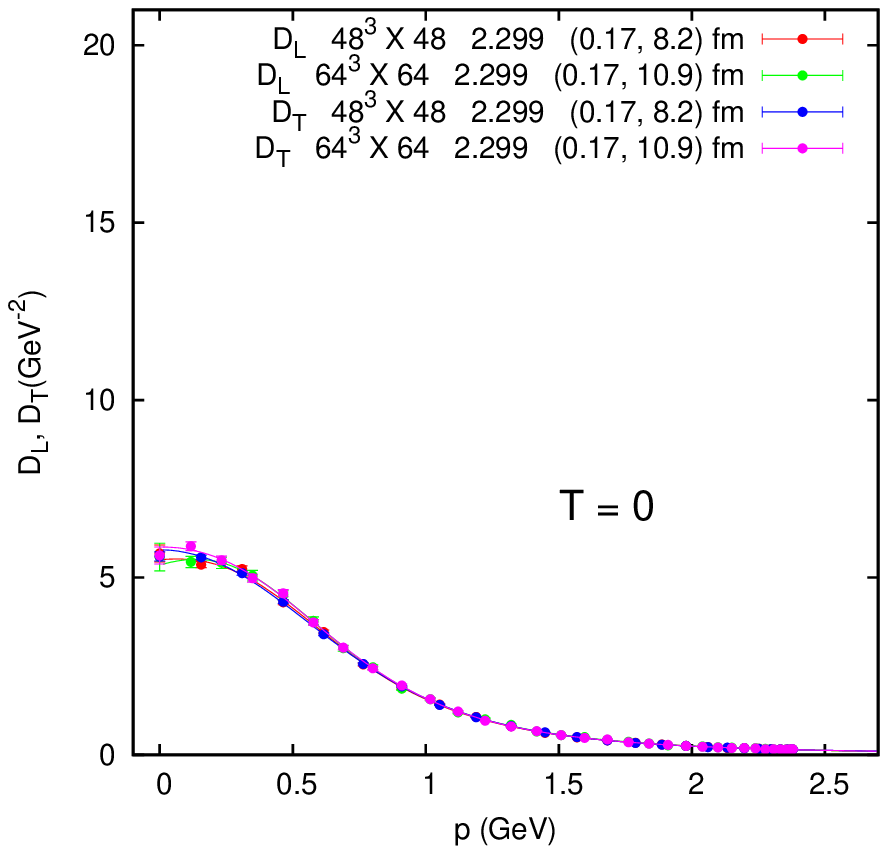}
\hspace*{-2.7cm}
\includegraphics[height=7.truecm]{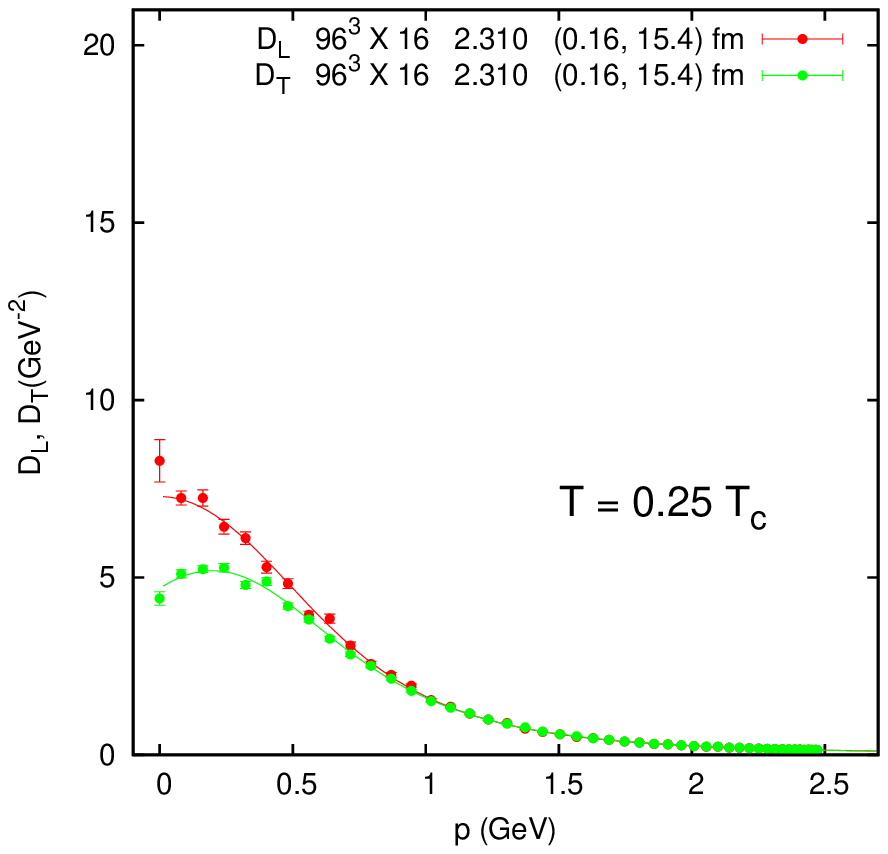}
\vspace*{2mm}
\hspace*{-1.5cm}
\includegraphics[height=7.truecm]{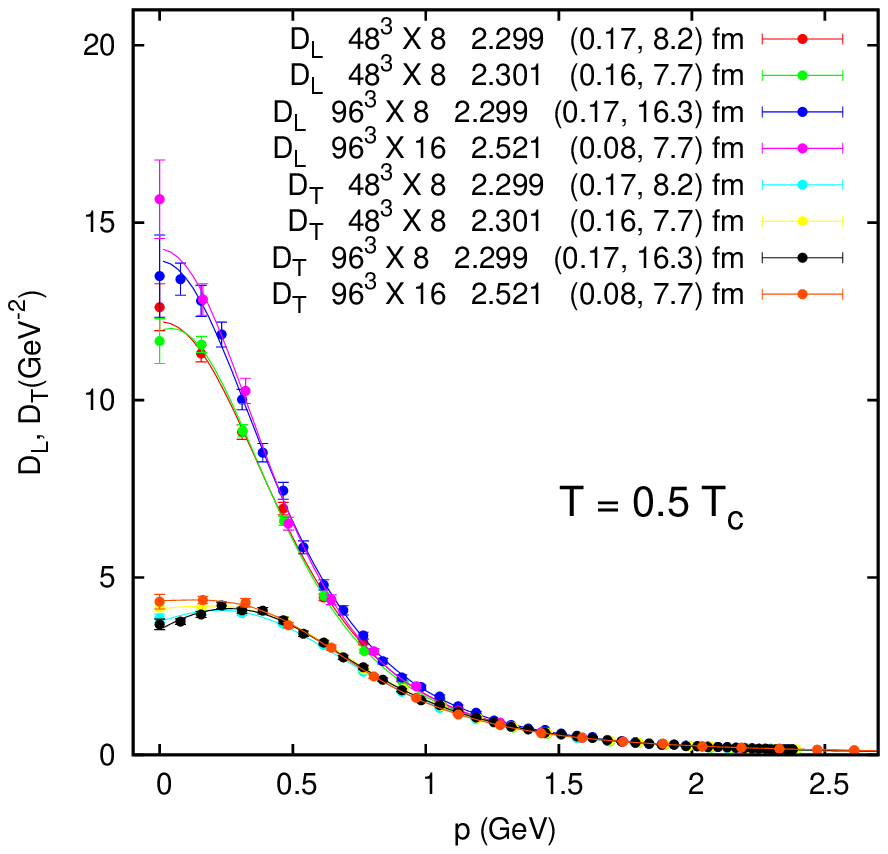}
\hspace*{-2.7cm}
\includegraphics[height=7.truecm]{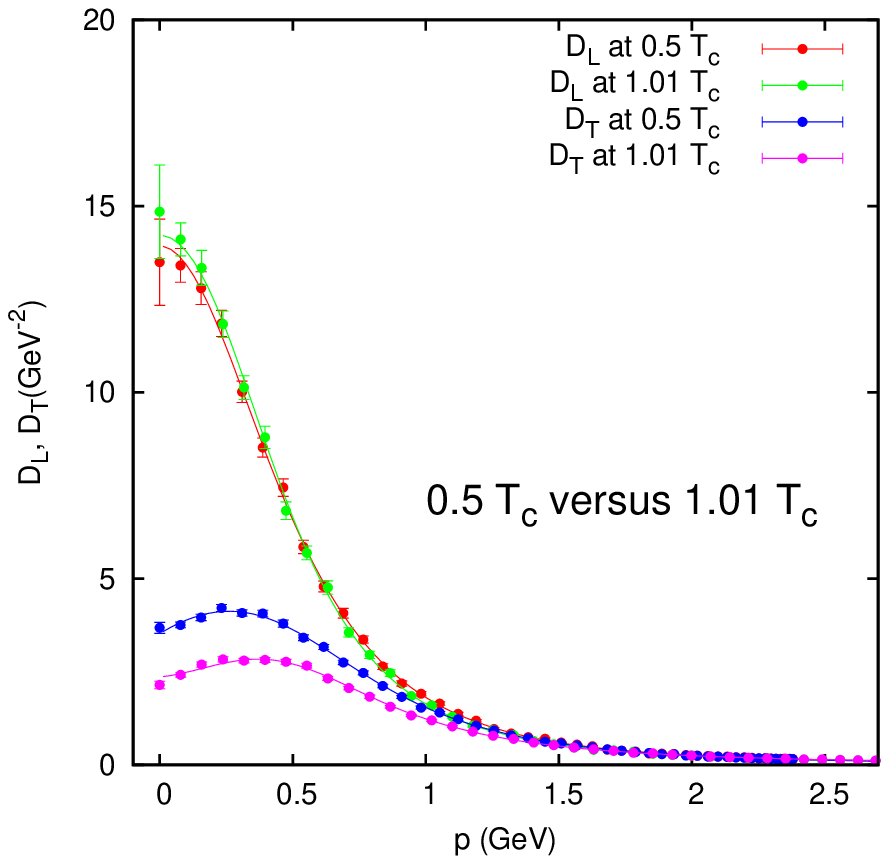}
\caption{Longitudinal and transverse gluon propagators at $T=0$ (top left),
$T=0.25 T_c$ (top right) and $T=0.5T_c$ (bottom left). Curves for
$T=0.5T_c$ and 1.01 $T_c$ are shown together for comparison on the
bottom right.
Values for the lattice volume $N_s^3 \times N_t$, the lattice parameter
$\,\beta$, the lattice spacing $a$ and spatial
lattice size $L$ (both in fm, in parentheses) are given in the plot
labels, with the exception of the bottom right plot, which is described
in the text.}
\label{combined}
\end{figure}

We show our fits, together with the data, for several values of
the temperature $T$ (given in terms of the critical temperature
$T_c$) in Fig.\ \ref{combined}. We see that $D_L(p^2)$ increases as 
the temperature is switched on, while $D_T(p^2)$ decreases slightly, 
showing a clear turnover point at around 350 MeV.
It is interesting to notice that the infrared behavior
of $D_L(p^2)$ remains unchanged (within errors) from $0.5 T_c$ to $T_c$,
as shown in the bottom right plot in the figure. (The curves shown
are for lattice parameters $\beta= 2.299, 2.515$ and lattice volumes 
$96^3 \times 8$, $192^3 \times 8$ respectively for the temperatures 
0.5 $T_c$ and 1.01 $T_c$.)
In fact, after reducing the severe systematic effects that are 
observed around $T_c$, we find a relatively smooth behavior of $D_L(p^2)$ 
with $T$, which calls into question the sensitivity of 
the electric propagator to the deconfinement transition.

%%%%%%%%%%%%%%%%%%%%%%%%%%%%%%%%%%%%%%%%%%%%%%%%%%%%%%%%%%%%%%%%%%%%%%%%%%%%

\section{Conclusions}
\label{conclusions}

By fitting rational functions of $p^2$ to the whole range of our
(infrared) data for the SU(2) Landau-gauge gluon propagator $D(p^2)$ in
four and three space-time dimensions, we are able to obtain estimates 
for the physical values of the masses in the RGZ framework, 
as well as to gain a better understanding of the pole structure in the 
proposed expressions.
The data points range from about 4 GeV down to 20--40 MeV, which are the
smallest simulated momenta to date.
In each case, we look for the best fit to the data, with the smallest number
of independent parameters, and relate them to the condensates in the proposed
analytic forms only at the end.
Put differently, the predicted dependence of the fit parameters on the
condensates is {\em not} imposed in the fitting form, but is obtained as
a result of the fit.

We find that the 4d results are well described by the simplified
version of the RGZ gluon propagator in Eq.\ (\ref{prop}), equivalent to the
simplest Gribov-Stingl form. This corresponds to a pair of complex-conjugate
poles, as opposed to the Gribov propagator, in which the poles would be
purely imaginary. 
The values for the condensates $M^2+\rho_1$, $m^2$ and $\lambda^4$
are in agreement with the ones obtained for the SU(3) case in
Ref.\ \cite{Dudal:2010tf}.
The quantitative agreement between the infrared limit of SU(2) and SU(3)
theories was observed numerically before in
\cite{Sternbeck:2007ug,Cucchieri:2007zm}.

In 3d, our fits support the more general form of the RGZ propagator
in Eq.\ (\ref{refinedgluonprop}).
In this case, the condensate $\rho$ is a complex quantity and there
are significant differences in the values of the other condensates
and of $\lambda^4$ compared to the 4d case. Also, in 3d one has a real
pole mass in addition to the pair of complex-conjugate poles.
It is interesting to note that the masses from the complex poles assume
similar values in 3d and 4d,
with an imaginary part that is more than twice their real part.
(We recall that a Gribov propagator would have a null real part.)
Note also that the mass and the coefficient
associated with the real pole in 3d are very small.

\vskip 3mm
Our analysis strongly suggests a pole structure
with complex-conjugate masses (with comparable real and imaginary parts)
for the infrared gluon propagator in Landau gauge, for zero-temperature
(in 4d and 3d) and for nonzero temperatures below and around the critical
temperature $T_c$.
As stressed at the end of Section \ref{RGZ}, one can
interpret this result as describing an unstable particle.
In particular (see \cite{Cucchieri:2011ig}), in the zero-temperature
4d case we obtain the values $m_g \approx 550$ MeV and
$\Gamma_g \approx 1180$ MeV respectively for the gluon mass and for
its width. The very large value for the width $\Gamma_g$ may be associated
to a lifetime $\tau_g$ smaller than 10$^{-24}$ s, supporting the existence 
of very short-lived excitations of the gluon field.

%%%%%%%%%%%%%%%%%%%%%%%%%%%%%%%%%%%%%%%%%%%%%%%%%%%%%%%%%%%%%%%%%%%%%%%%%%%%

\section*{Acknowledgments}

D.~Dudal and N.~Vandersickel are supported by the Research-Foundation
Flanders (FWO). A.~Cucchieri and T.~Mendes thank CNPq and FAPESP for
partial support. A.~Cucchieri also acknowledges financial support from
the Special Research Fund of Ghent University (BOF UGent).

%%%%%%%%%%%%%%%%%%%%%%%%%%%%%%%%%%%%%%%%%%%%%%%%%%%%%%%%%%%%%%%%%%%%%%%%%%%%

\end{document}